\documentclass[runningheads]{llncs}
\usepackage[T1]{fontenc}
\usepackage{amsmath,amssymb,amsfonts}
\usepackage{graphicx}
\usepackage{float}
\usepackage{subcaption}
\usepackage[export]{adjustbox}
\usepackage{placeins} 
\usepackage{fancyhdr}
\usepackage{lipsum} 
\begin{document}

\title{Multi-Scale Convolutional LSTM with Transfer Learning for Anomaly Detection in Cellular Networks}
\pagestyle{fancy}
\fancyhf{} 
\fancyhead[RO,LE]{\thepage} 
\fancyhead[LO]{\leftmark} 
\fancyhead[RE]{TL for Anomaly Detection in Cellular Networks} 

\renewcommand{\headrulewidth}{0pt} 
\renewcommand{\footrulewidth}{0pt} 

%
\author{Nooruddin Noonari\inst{1}\orcidID{0000-0002-3843-0415} \and
Daniel Corujo\inst{2}\orcidID{0000-0002-7484-1027} \and
Rui L. Aguiar\inst{3}\orcidID{0000-0003-0107-6253}
\and
Francisco J. Ferrão\inst{4}\orcidID{0009-0004-6410-2468}
}
\authorrunning{Nooruddin et al.}
%
\institute{Instituto de Telecomunicações in the Universidade de
Aveiro, 3810-164 Aveiro, Portugal
\\
\email{nooruddin@ua.pt,dcorujo@ua.pt,francisco.ferrao@datamachineelite.com}\\
\and Instituto de Telecomunicações in the Universidade de Aveiro \and Instituto de Telecomunicações in the Universidade de Aveiro \and Data Machine Elite, Alenquer, Portugal
\\
}
\maketitle              
\begin{abstract}
The rapid growth in mobile broadband usage and the increase in subscribers make it crucial to ensure these networks are reliable and perform well. As more people use mobile networks daily, these networks become increasingly complex, especially during peak hours. Manually collecting reports of Key Performance Indicators (KPIs) to understand network performance takes a lot of time due to the vast amount of data. Quickly detecting network failures and identifying unusual behavior during busy times is crucial for determining if the network is functioning normally. Researchers have used Deep Learning (DL) and Machine Learning (ML) techniques to understand how networks behave. These techniques include predicting data throughput, analyzing call records, and detecting when a cell is out of service. However, these methods need much computational power to train from scratch and predict network behavior. They also need large amounts of labeled data and are usually specialized for specific scenarios, requiring retraining for new applications. Training from scratch is time-consuming and costly because it needs enough training data and computational resources.

This study introduces a new supervised learning method, Multi-Scale Convolutional LSTM with Transfer Learning (TL), to detect unusual activity in cellular networks. First, the model is trained from scratch using a publicly available dataset to learn how cellular networks typically behave. Then, using the Transfer Learning approach, the trained Multi-Scale Convolutional LSTM model is fine-tuned by transferring weights and applying them to different datasets to detect anomalies. The study compares the model's performance from scratch with the fine-tuned Multi-Scale Convolutional LSTM model that uses the TL weight transfer technique. Additionally, Exploratory Data Analysis (EDA) and the Synthetic Minority Over-sampling Technique (SMOTE) address the class imbalance and better understand the dataset. The results show that our model, when trained from scratch, achieves an accuracy of 99\% after 100 epochs. The same fine-tuned model applied to a different dataset achieves 95\% accuracy after just 20 epochs. 
\end{abstract}
\section{Introduction}
With the widespread utilization of mobile cellular networks, such as 4G and now 5G \cite{b25}, people's access to internet services has become more accessible and essential. The rapid growth in mobile broadband usage and the increase in subscribers have made it crucial to ensure the reliability and performance of these networks. The importance of throughput in mobile cellular networks has increased drastically throughout the evolution of mobile networks. Ensuring connectivity, reliability, Quality of Service (QoS), and overall performance is essential. The daily increase in subscribers has led to the complexity of mobile cellular networks, making it even more critical to ensure uninterrupted service, particularly during peak hours\cite{b1,b2,b3}. KPIs measure network performance and are regularly collected from various cells. Manually detecting these network failures is time-consuming due to the vast amount of data, with hundreds of KPIs. It is essential to detect network failures effectively and quickly. Identifying unusual behavior in cellular networks during peak hours is crucial; in other words, detecting anomalies is critical to determine if the network behavior is normal or abnormal during these times\cite{b4,b5}.

Researchers have employed various methodologies to understand network behavior, including throughput prediction \cite{b6}, call records analysis \cite{b7}, and cell outage detection \cite{b8}. To enhance the accuracy of these predictions, several DL and ML techniques have been utilized, such as LSTM+1D \cite{b9}, general ML approaches \cite{b10}, and an Automatic Labeling Technique for Supervised Learning \cite{b11}. Consequently, these DL techniques, such as Convolutional Neural Networks (CNNs) and Long Short-Term Memory (LSTM), require computational resources for training from scratch and subsequent prediction of network behavior. These methods have limitations, such as the need for large amounts of labeled data, and traditional methodologies are typically specialized for specific scenarios and require further retraining when applied to new wireless network applications. Unfortunately, training from scratch requires computation resources and enough training data, which is a time-consuming and costly difficult task\cite{b12,b13}. Recently, DL introduced a technique called TL, which helps resolve dataset labeling and computational problems. In the TL technique, a DL model can be trained on one application and reused for a different but related task\cite{b14}.

This study proposes a supervised learning approach,  Multi-Scale Convolutional LSTM with TL for Anomaly detection in cellular networks. This approach is initially trained from scratch on a publicly available dataset to comprehend cellular network behavior. Additionally, this study utilizes weight transfer by first training the model from scratch for anomaly detection and subsequently applying it to different datasets using a TL approach. 

The article is structured as follows: Section 2 provides an overview of related past work in the literature review. Section 3 discusses the methodology. Section 4 presents the results with a comprehensive analysis of the findings. Finally, Section 5 summarizes the work.

\section{State of the Art}
Recently several DL and ML-based approaches have been explored to detect anomalies in various fields. 

Researchers tried to implement \cite{b15} Recurrent Neural Networks (RNN) for computing DL-Throughput to understand the performance of networks. Such a technique reduced the prediction error by 29\% compared to traditional prediction techniques. Furthermore, an improved version of RNN and LSTM \cite{b16} was implemented to detect the DL-Throughput in LTE networks. They have utilized real-time approaches such as video streaming to understand the network performance. Another study \cite{b9} uses LSTM and 1D CNN for detection and forecasting network behavior through densely deployed femtocell outage detection for cellular networks. The approach implemented aggregation decision methods with a combination of LSTM and 1D CNN to understand the reason for cell outages, As a result, it calculated the cell outage downtime using feed-forward neural networks (FFNN) with ultra-dense cellular networks. Another approach \cite{b17} implemented automatic network traffic anomaly detection using an ML approach. The research paper addressed both detection and classification, focusing specifically on traffic anomalies in cellular networks. The methodologies implemented include decision trees, SVM, and neural networks. But all these models require training from scratch which requires substantial computational resources and sufficient training data, making it a time-consuming and costly endeavor.

To address dataset labeling issues and facilitate training from scratch, researchers have recently applied Transfer Learning across several research areas, including time series, such as \cite{b18} which proposed a TL-based fault diagnosis method using nearest neighbor feature constraints to improve feature extraction and enhance domain adaptation performance. Another researcher \cite{b19} utilizes TSTL to transfer early-stage time series to serious stages, enhancing fault feature detection and improving model adaptability to varied conditions. Transfer learning has proven to be a valuable technique for improving the performance and efficiency of anomaly detection and fault diagnosis in time series data, especially within industrial settings. For instance, the MENDEL \cite{b20} framework demonstrates an effective application of transfer learning to detect anomalies in industrial control systems. By leveraging pre-trained models and using Principal Components Analysis (PCA) to reduce features, MENDEL achieves high performance with minimal training data. This approach ensures that models can be effectively transferred and adapted across different industrial domains, showcasing significant improvements in detection accuracy with a reduced amount of training data. Despite the extensive application of TL in various research fields, there remains significant potential for further investigation. For instance, the use of TL in cellular networks still presents gaps and offers opportunities for continued research.

\section{Research Methodology}
This study proposes a supervised learning approach, Multi-Scale Convolutional LSTM with TL, for anomaly detection in cellular networks. 

\subsection{Convolutional Neural Network (CNN)}

A CNN is a type of Deep Neural Network (DNN) that is primarily used for image recognition and Computer Vision (CV) tasks. CNNs are designed to automatically and adaptively learn spatial hierarchies of features from input data. CNN consists of fully connected layers, an input layer, an output layer, and several intermediate layers known as hidden layers and each neuron is connected to every neuron in the previous and the next layer. Neurons compute values using the weighted sum of inputs and a bias term, with activation functions enabling non-linear learning.

\subsection{Long Short-Term Memory (LSTM)}

An LSTM network, a type of Recurrent Neural Network (RNN), addresses the vanishing and exploding gradient problem common in RNNs and uses memory cells and gates to manage information flow, categorical features are one-hot encoded into binary vectors, which are then concatenated and fed into the LSTM's input layer. The network's hidden layer comprises memory cells and three types of gates: input, forget, and output. The input gate controls what new information enters the memory cell, the forget gate determines which previous information to discard, and the output gate decides what information to pass on from the memory cell.

\subsection{Multi-Scale Convolutional LSTM}

\begin{figure}[H]
    \centering
    \includegraphics[width=1.0\linewidth]{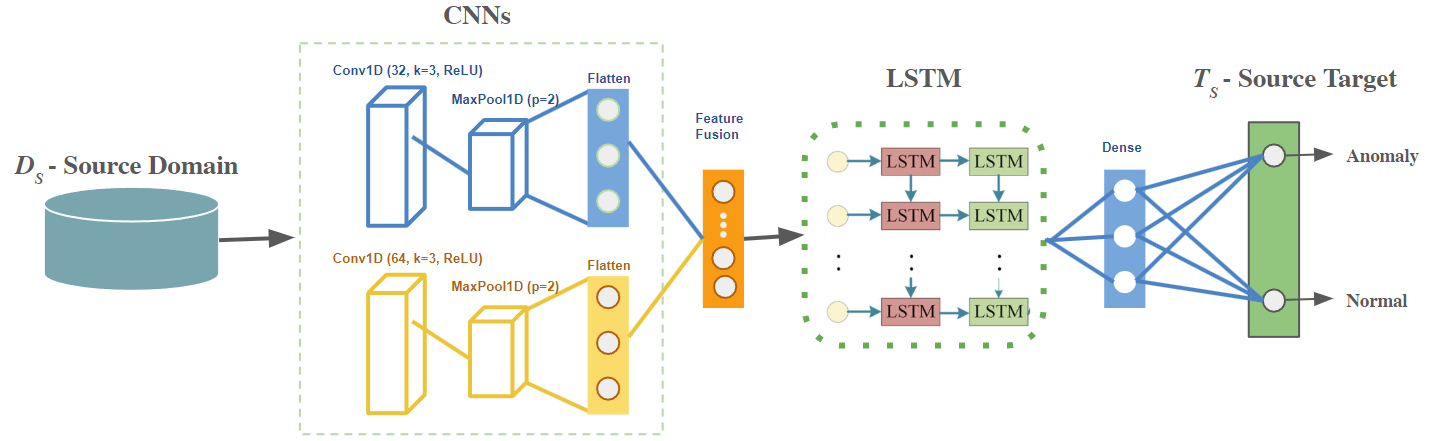}
    \caption{Our framework Multi-Scale Convolutional LSTM.}
    \label{fig:MSC-LSTM}
\end{figure}

As shown in the Fig.~\ref{fig:MSC-LSTM}, In this study, the proposed Multi-Scale Convolutional LSTM model combines CNN and LSTM layers to capture spatial and temporal features. The input data shape includes one feature dimension. Two CNN block layers extract features: the first with 32 filters and the second with 64 filters, both using ReLU activation, MaxPooling, and flattening. These outputs are fused into a single vector. Two LSTM layers handle temporal dependencies: the first returns sequences, and the second returns the final output. A Dense layer with 100 units and ReLU activation process data, followed by an output layer with a sigmoid activation function, predicting anomalies between 0 (normal) and 1 (anomaly).

Inspired by \cite{b27}, we have designed a model with fewer convolutional layers and trained our model for 100 epochs compared to 400 epochs as the original model.

\subsection{Introduction to Transfer Learning (TL)}

TL is an ML technique where a model trained for one task is reused for another related task. In TL, the knowledge gained from solving one problem is applied to a different but related problem. TL offers a promising approach to address these challenges. TL involves two fundamental components: domain and task \cite{b24}.

\begin{figure}[H]
    \centering
    \includegraphics[width=1.0\linewidth]{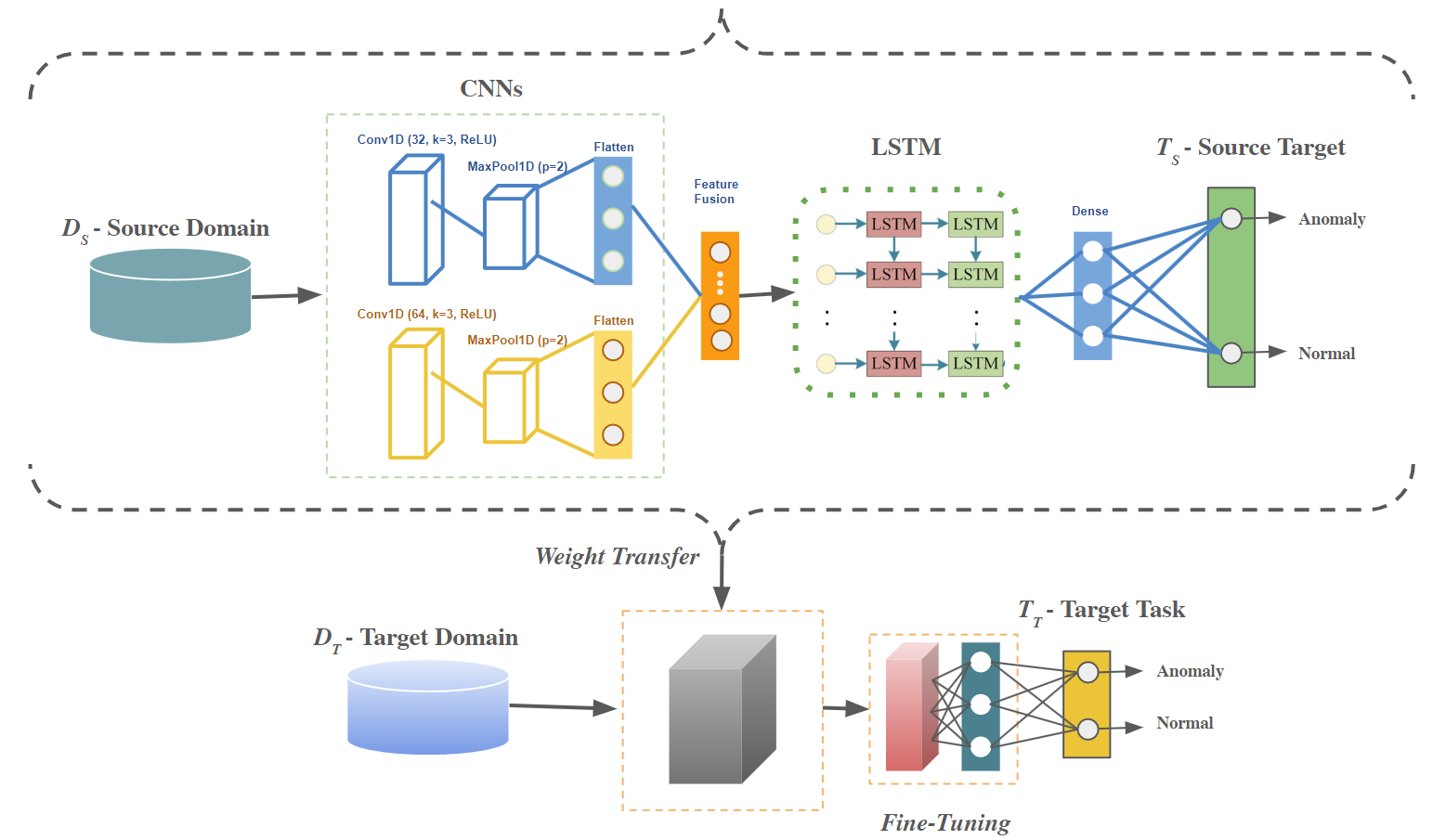}
    \caption{TL for Anomaly Detection in Cellular Networks.}
    \label{fig:TLapproach}
\end{figure}

In the context of TL, a domain \(D = \{X, P(x)\}\) consists of a feature space \(X\) and a marginal probability distribution \(P(x)\). Here, \(x = \{x_1, x_2, \ldots, x_i\}\) represents feature values, with \(x_i\) belonging to \(X\) and \(i\) indicating the number of samples. A task \(T = \{L, \theta\}\) includes a label space \(L\) and parameters \(\theta\) of a function \(f(\cdot)\) that is learned during training, using data pairs \(\{x_i, y_i\}\), where \(x_i \in X\) and \(y_i \in Y\). Alternatively, the task can be defined as \(T = \{y, P(y|x)\}\), where \(y\) represents the labels and \(P(y|x)\) is the conditional probability. TL leverages a source domain \(D_{S}\) and task \(T_{S}\) to improve learning in a target domain \(D_{T}\). The pre-trained parameters \(\theta\) from the source function \(f(\cdot)\) are used to learn the target distribution \(P(y_{T}|x_{T})\), with the conditions that \(D_{S} \neq D_{T}\) and \(T_{S} \neq T_{T}\) as shown Fig.~\ref{fig:TLapproach}.

\subsection{Our Approach}
In our approach, we proposed two strategies: We designed the Multi-Scale Convolutional LSTM model and performed the initial training on one dataset. We fine-tuned that model as a pre-trained model and used it for the TL approach. The Multi-Scale Convolutional LSTM model consisted of two CNNs for feature fusion. We used those features with LSTM for anomaly detection in a cellular network and saved its weights. In another task, we used these weights to fine-tune the model for anomaly detection in a cellular network with another dataset to reduce the computation resources.

\section{Experimental Results}
\subsection{Pre-Processing and Experimental Setup}
The experiment was conducted on a CPU with 36 GB RAM and GPU configured as Nvidia RTX 3060. The Keras library and other essential Python libraries were employed to implement the Multi-Scale Convolutional LSTM model.

\subsubsection{Dataset Collection}
The dataset consists of two measurements of cellular networks collected from Kaggle, one called Dataset (1920) \cite{b28} and the second called Dataset (2021) \cite{b29}. We have used Dataset (1920) for Multi-Scale Convolutional LSTM for training from scratch and  Dataset (2021) for fine-tuning models. The dataset is derived from a real LTE network. Over two weeks, researchers collected various measurements from 10 base stations every 15 minutes. Each base station has a different number of cells. The data is stored in a CSV file, with each row representing a sample from one cell at a specific time. Each sample includes several features.

\subsubsection{Performance Metrics}
To comprehensively evaluate the results, we employed two primary metrics: the loss function and various evaluation metrics. We used Binary Cross-Entropy (BCE) as the loss function for anomaly detection in cellular networks. For further results evaluation, we calculated accuracy, precision, recall, and F1-score to assess the performance of the Multi-Scale Convolutional LSTM model.
\begin{figure}
    \centering
    \includegraphics[width=0.8\linewidth]{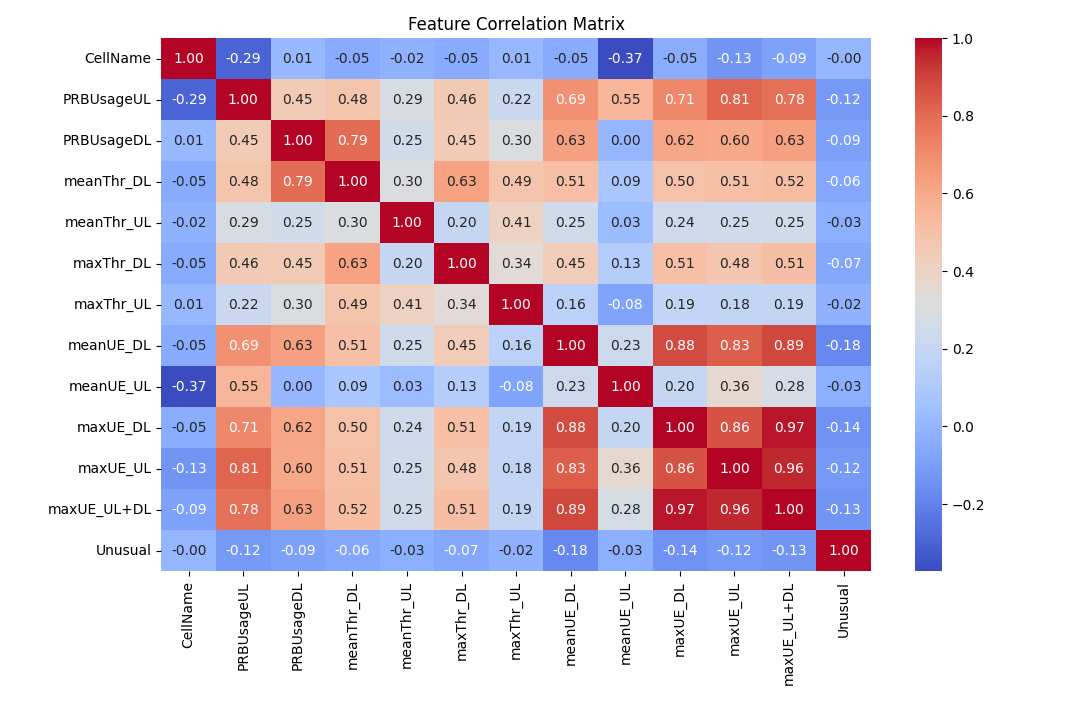}
    \caption{FCA Results.}
    \label{fig:FeatureCoMa}
\end{figure}

\begin{figure}
    \centering
    \includegraphics[width=0.8\linewidth]{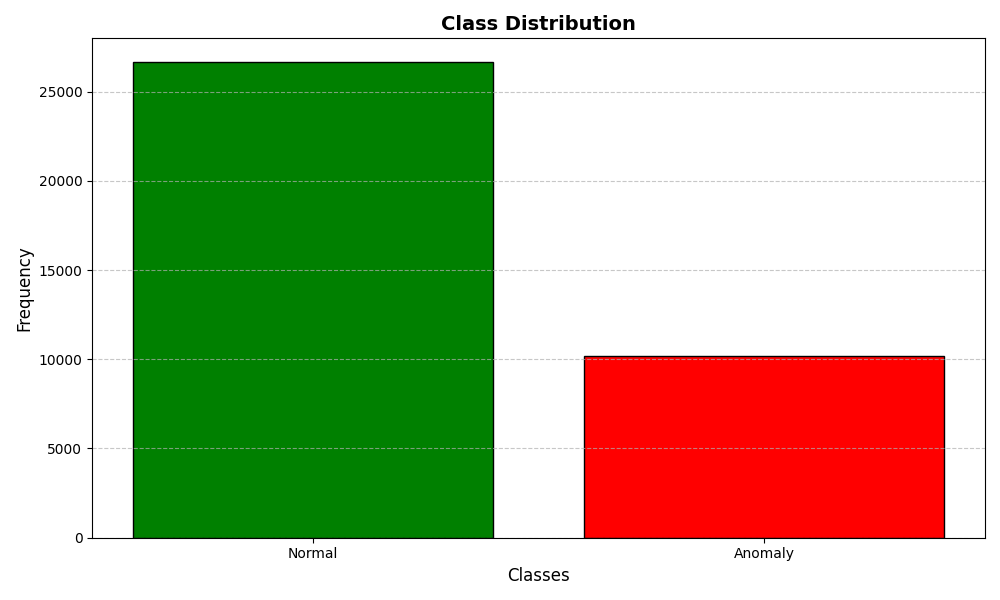}
    \caption{Class Imbalance Problem Results}
    \label{fig:classImPr}
\end{figure}
\subsection{Exploratory Data Analysis (EDA) Results}

EDA is the initial step in examining a dataset to understand its main features and identify patterns, unusual data points, and connections between variables. It allows data scientists and analysts to make informed decisions and discover insights. For our task, we used various EDA techniques, such as Feature Correlation and SMOTE as class imbalance techniques, to analyze the cellular network dataset. Our dataset has highly imbalanced classes, with 80\% normal data (0) and 20\% anomalies (1), as shown in Fig.~\ref{fig:classImPr}. This imbalance made achieving high validation accuracy challenging, so we applied the SMOTE technique to balance the dataset. In  Fig.~\ref{fig:FeatureCoMa}, each cell represents the correlation between two features: red indicates a strong positive correlation, blue is a strong negative correlation, and white has no significant relationship.

\subsection{Multi-Scale Convolutional LSTM Results}

\begin{figure}
    \centering
    \begin{minipage}[b]{0.45\textwidth}
        \centering
        \includegraphics[width=1.0\linewidth]{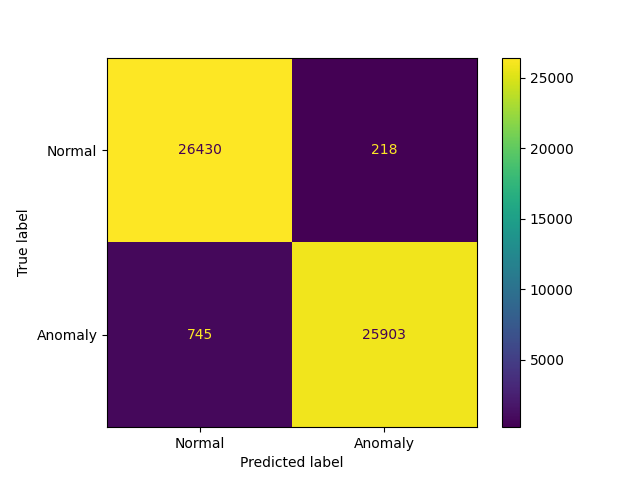}
        \subcaption{MSC-LSTM Confusion Matrix Results. }\label{fig:Result_1}
    \end{minipage}
    \hspace{0.05\textwidth}
    \begin{minipage}[b]{1.0\textwidth}
        \centering
        \includegraphics[width=1.0\linewidth]{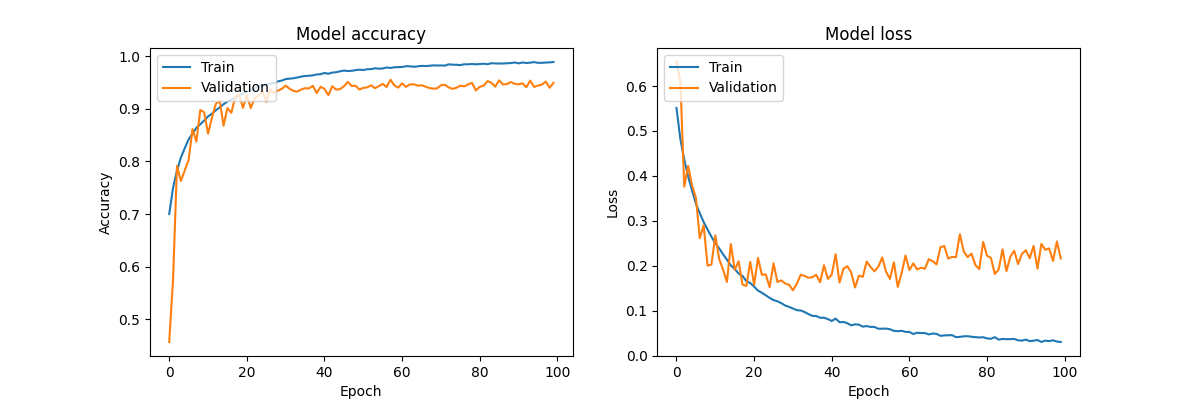}
        \subcaption{MSC-LSTM Accuracy and Loss.}\label{fig:Result_2}
    \end{minipage}
    \caption{Model Performance Metrics}\label{fig:Results}
\end{figure}

In our methodology, we conducted supervised learning, using the last feature labeled 0 for normal and 1 for anomaly to understand the behavior of cellular networks. Fig.~\ref{fig:Result_2} shows the results of training and validation accuracy, as well as training and validation loss. The prediction accuracy for anomaly detection achieved by our model is 99\% over 100 epochs, which is significantly better than existing models. 

Table~\ref{tab:01} presents the performance metrics for each label including Normal, and Anomaly, derived from the classification report after training and prediction. For the "Normal" label in cellular networks, the precision is 97\%, recall is 100\%, and the F1-score is 99\%. For the "Anomaly" label, the precision is 100\%, recall is 97\%, and the F1-score is 99\%, indicating excellent performance in detecting cellular network behavior.

\begin{table}
\centering
\caption{MSC-LSTM performance Evaluation.}
\begin{tabular}{|c|c|c|c|}
\hline
\textbf{Labels} & \textbf{Precision (\%)} & \textbf{Recall (\%)} & \textbf{F1-score (\%)} \\
\hline
Normal & 97\% & 100\% & 99\% \\
\hline
Anomaly & 100\% & 97\% & 98\% \\
\hline
Accuracy & 99\% & Epochs & 100 \\
\hline
\end{tabular}

\label{tab:01}
\end{table}

Furthermore, Fig.~\ref{fig:Result_1} includes the confusion matrix results, providing a comprehensive analysis of anomaly detection and classification. The model distinguishes between "Normal" and "Anomaly" cases effectively. The matrix shows that the model correctly identified 26,430 instances as "Normal" and 25,903 instances as "Anomaly." However, there were 745 "Anomaly" instances incorrectly labelled as "Normal" and 218 "Normal" instances misclassified as "Anomaly." Brighter colours in the matrix indicate higher prediction counts, enabling a quick visual assessment of model accuracy and error distribution.

\begin{figure} [H]
    \centering
     \includegraphics[width=1.0\linewidth]{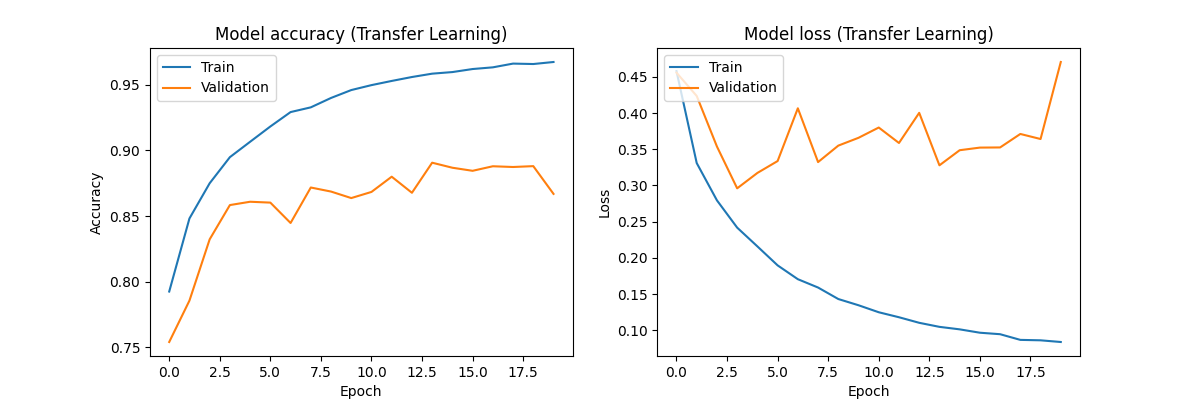}
    \caption{Fine-tune Model Accuracy and Loss.}
    \label{fig:TL_Result_2}
\end{figure}

\subsection{TL based fine-tuned Results}

Furthermore, in our study, we fine-tuned and trained the model for 20 epochs using a Dataset(2021) to detect unusual patterns in cellular networks.
Fig.~\ref{fig:TL_Result_2} displays the results and shows that our fine-tuned model achieved a far better accuracy with fewer iterations, which states that the TL approach helps to detect anomalies in cellular networks and reduce computation power.

\section{Conclusion}
In this study, we focused on improving the detection of unusual activity in cellular networks. We introduced a new supervised learning method called Multi-Scale Convolutional LSTM. We performed three tasks: first, we performed EDA techniques to preprocess the dataset, and then we used our model to train from scratch and saved its weights for use in the TL approach. Furthermore, our model is initially trained using a publicly available dataset to understand normal cellular network behavior. Through TL, we fine-tune the model by transferring weights and applying them to different datasets to detect anomalies more effectively. We addressed the challenges of class imbalance. Our results demonstrated that when trained from scratch, the Multi-Scale Convolutional LSTM model achieved a high accuracy of 99\% after 100 epochs. Furthermore, when we fine-tuned the model using TL, it achieved an impressive 95\%  accuracy after only 20 epochs on a different dataset.

Considering future work, we will extend this for fault detection and compare it with different DL/ML approaches.

\section{Acknowledgement}
This work was supported by the European Union’s Horizon Europe Marie Skłodowska Curie Action SCION under No. 101072375.

\end{document}